# Subwave length light focusing by surface plasmons on a silver film

V. S. Zuev and G. Ya. Zueva

The P. N. Lebedev Physical Institute of RAS
53 Leninsky pr., 119991 Moscow Russia

Two devices for subwave length focusing of light are explored. The first one is a thin film of a well reflecting metal which the converging beam of surface plasmons with a wave number $h >> \omega_0 / c$ is excited on. The waist of this beam has a cross section that is much smaller than $\lambda_0^2$. The second device is a linear nanoantenna with a gap in the middle. A resonance plasmon is excited in the antenna. The field in the gap concentrates on the spot that is much smaller than $\lambda_0^2$. In both cases the effect of field enhancement in the spot of subwave length focusing is greatly decreased because of propagation losses in the first case and because of a small excitation cross section in the second case. It is suggested to improve on the effect of excitation of the film and the nanoantenna by means of interaction with an adjacent atom or quantum dot.

There are two recent proposals on how to focus light into a spot that would be much smaller than a wave length. The first proposal could be found in the Internet as titles of lectures that the famous Professor E. Yablonovitch delivered in 2006 at some meetings. The titles are "Plasmonics: Optical Frequencies but with X-ray Wavelengths" and "What is the Limit of Focusing Light ?" /2/. The mentioned lectures are accessible by the titles and abstracts only. Unfortunately the abstracts contain a poor quantity of information. The substantial information could be found in a photo that is presented at the site of The 89th OSA Annual Meeting Frontiers in Optics 2005 & Laser Science XXI Conference, October 16-20, 2005. The reconstruction of the photo is on the Fig. 1 below. It is seen that it is a device with a surface plasmons that the author would keep in mind.

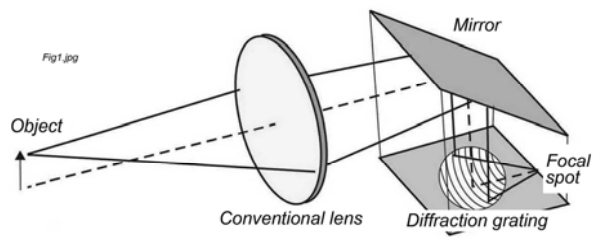

Fig. 1.

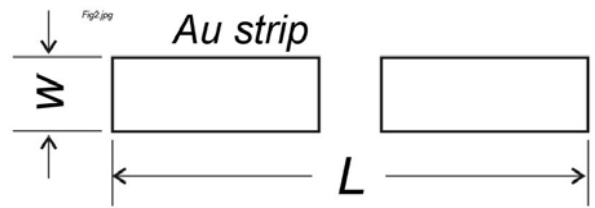

Fig. 2.

The second proposal could be found in the paper by Professor D. W. Pohl and coauthors /3/. They demonstrated the light localization on a spot of $20\ nm \times 45\ nm$ by a device that is called a resonant optical antenna. A sketch of the antenna is shown in Fig. 2. It is an Au small thin strip with a gap in the middle on a glass substrate. The field in the gap is greatly enhanced.

In the Yablonovitch's device it is a light beam that is focused on a uniform thin metal film. In the focus spot of the $\sim \lambda_0$ diameter there is a diffraction grating that consists of a set of cuts of concentric circles. The diffraction grating excites a converging beam of surface plasmons in the film. The plasmons focalize into a spot which dimension is much smaller than the dimension of the initial focusing spot.

The Pohl's device is an Au strip 40 nm thick $\sim$ 260 nm long and 45 nm wide with a gap 20 nm wide in the middle. The strip was placed in a focal spot of a conventional lens. A conversion of the laser light (830 nm pulses of a duration 8 $ps$) into intense white light was observed. The mechanisms underlying the effect are not well known. However the authors are sure that the effect points unambiguously to a very high intensity of the field in the gap.

Let us discuss in details the peculiar features of the mentioned devices. In the Yablonovitch's device the key element is the film with a converging beam of surface plasmons. It is possible to excite an optical plasmon with a wave length as short as 10 nm on surfaces of Ag and Au films /4,5/. This plasmon has the wave length many times as short as compared to a wave in vacuum of the same frequency. Great additional shortening of the plasmon wave length could be accomplished if the metal film would be covered by a dielectric layer of $\varepsilon$ that is lower but close to the module $\varepsilon$ of the metal /5/.

Au films and very probably Ag films as thin as 0.5 nm without an island structure can be produced by the existent contemporary technologies /6/. Though small surface inhomogeneities should not produce any difficulty because they emit radiation poorly.

There are various means for describing a focusing phenomenon. It is convenient to do this by a concept of a Gaussian beam. A Gaussian beam arises as a solution of the Helmholtz equation in the paraxial approximation /7,8/. On a flat surface separating two substances surface plasmons present by themselves plain waves however so called nonuniform or evanescent waves. They are of a variable amplitude along the wave front. Besides the plasmon field has different descriptions on two sides of the separating surface. The description of a Gaussian beam of surface plasmons was not found as yet by us in the special literature.

Below it will be a so called $TM$ - plasmon under a consideration. This plasmon has a single magnetic component, namely a transverse $H_x$ component parallel to the separating surface, see Fig. 3. The electric field has two components: a transverse $E_y$ component that is a normal to the separating surface and a longitudinal $E_z$ component parallel to the separating surface. There are two such plasmons symmetric (s-plasmon) and antisimmetric (a-plasmon) on a thin film. The highly slowed plasmon that is of a short wave length is an a-plasmon. It is an a-plasmon that will be dealed with below.

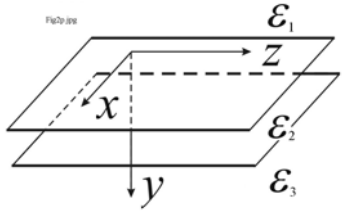

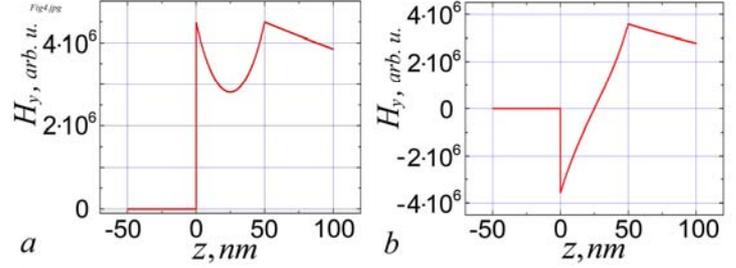

Fig. 3.                                    Fig. 4.

The following equation is called the Helmholtz equation:

$$\Delta \Phi + (\omega / c)^2 \varepsilon \mu \Phi = 0 . \qquad (1)$$

It arrives from the Maxwell equations for a free monochromatic field of radiation of the frequency $\omega$. In the Descartes coordinates $\Delta = (\partial^2 / \partial x^2 + \partial^2 / \partial y^2 + \partial^2 / \partial z^2)$ and the equation (1) is applicable to each component of each field vector – a vector-potential $\vec{A}$, an electrical strength $\vec{E}$ and a displacement $\vec{D}$, a magnetic induction $\vec{B}$ and a field $\vec{H}$.

Two planes in the Fig. 3 present surfaces of a metal film with a dielectric constant $\varepsilon_2 < 0$. Adjacent media have dielectric constants $\varepsilon_1$ and $\varepsilon_3$, $|\varepsilon_2| > \varepsilon_1, \varepsilon_3$. The $x$ and $z$ axes lie in the plain of media separation the $y$ axis is a normal to the separating plain. The coordinate system origin lies on the separating plain. A surface wave propagates along the $z$ axis and does not depend on $x$.

A plasmon on a thin film consists of four plain evanescent waves of the appearance

$$H_{ix} = H_{ix} e^{q_{ij} y} e^{-ihz} , \ i = 1,2,3, \ j = 1,2 . \qquad (2)$$

Here $i$ is a medium index and $j$ is a number of a wave. In the $i = 1$ medium there is a single $j = 1$ wave that vanishes on moving toward $y \to -\infty$, in the $i = 3$ medium there is a single $j = 4$ wave that vanishes toward $y \to \infty$, in the medium $i = 2$ (in the metal) there are two waves. The $j = 2$ wave vanishes on moving in the positive direction along the $y$ axis and the $j = 3$ wave vanishes in the opposite direction. The wave amplitudes and a dispersion relation originate on equating the tangential components of the electric and magnetic fields on the plains separating substances. The magnetic fields for a s-plasmon $(a)$ and an a-plasmon $(b)$ on an $Ag$ film 50 nm thick in vacuum at the wave length $\sim 500$ nm are shown in Fig. 4.

All the four waves have the same $h$ value, for $q_{ij}$ the following relations take place

$$k_0^2 \varepsilon_i \mu_i = h^2 - q_{ij}^2 , \ k_0 = \omega / c . \qquad (3)$$

The $h$ value arises from the dispersion relation if the natural waves (eigenfunctions) of the film are considered. On excitation from the outside the $h$ can be of an arbitrary value /4/.

In order to receive the Helmholz equation for evanescent wave in paraxial approximation the $\Phi = V(x,y,z)e^{qy}e^{ihz}$ solution should be selected and put into (1). All the components of the considered $TM$-wave have the same coordinate dependence $e^{qy}e^{ihz}$ and differ only by numerical factors. As the result the next equation emerges:

$$(\partial^2/\partial x^2 + \partial^2/\partial y^2 + \partial^2/\partial z^2 + 2ih\partial/\partial z + 2q\partial/\partial y)V = 0 . \qquad (4)$$

Let cast aside $\partial^2 V/\partial y^2 + \partial^2 V/\partial z^2$ and receive

$$\partial V/\partial z - (iq/h)\partial V/\partial y = (i/2h)\partial^2 V/\partial x^2 . \qquad (5)$$

If the absence of the dependence of $V(x,y,z)$ on $y$ will be adopted then $\partial V/\partial y$ disappears and in result the next equation appears:

$$(\partial/\partial z)V(x,z) = (i/2h)(\partial^2/\partial x^2)V(x,z) . \qquad (6)$$

The equation (6) coincides with the equation for two-dimensional Gaussian beam /7/. The only difference is that $h >> (\omega/c)(\varepsilon\mu)^{1/2}$ instead of $(\omega/c)(\varepsilon\mu)^{1/2}$ appears. The solution of (6) is the function

$$V(x,z) = (hz)^{-1/2}e^{ihx^2/2z} . \qquad (7)$$

Let take a notice that (6) and (7) do not contain $q$ . As a result all four waves of the kind (2) that constitute a plasmon on a thin film are described by a single equation (7). By doing same the way as in /7/ the equation (7) has to be multiplied by the constant $\sqrt{-iha^2}$ and the displacement $z \to z - iha^2$ has to be done. In result the following expression for a Gaussian beam of surface plasmons appears:

$$u(x,y,z) = \sqrt{-i\frac{ha^2}{z - iha^2}} \exp\left(\frac{i}{2}\frac{hx^2}{z - iha^2} + q_{ij}y + ihz\right). \qquad (8)$$

The disregard of the $y$ and $z$ second derivatives leads to the condition $ha >> 1$ that resembles the common condition $ka >> 1$ for a Gaussian beam.

On having a concept of a Gaussian beam of plasmons the estimate $2a >> \lambda_{pl}/\pi$ for the minimal transverse dimension $2a$ of a converging beam of plasmons appears immediately. It is admissible to take $(2a)_{min} = 3\lambda_{pl}$ . The transverse dimension in the perpendicular direction is equal to $1/q_1 + 1/q_3 + t \approx 2/h + t = \lambda_{pl}/\pi + t$ , $t$ being the thickness of the film.

Let produce calculation for an $Ag$ film 1 nm thick at the wave length $\lambda_0 = 514.6\ nm$ . From the consideration performed in /5/ the following equation arises for an a-plasmon:

$$\left(\frac{q_2\varepsilon_1}{\varepsilon_2}\frac{1 + e^{-q_2d}}{1 - e^{-q_2d}}\right)^2 - q_2^2 = (\omega/c)^2(\varepsilon_2\mu_2 - \varepsilon_1\mu_1) . \qquad (9)$$

On putting $\varepsilon_{1,3} = 1$ , $\varepsilon_2 = -10.67$ /9/, $\mu_{i=1,2,3} = 1$ , $k_0 = \frac{2\pi}{\lambda_0} = 1.221 \cdot 10^5\ cm^{-1}$ ( $\lambda_0 = 514.6\ nm$ ) a resonance a-plasmon wave number $h = \sqrt{q_2^2 + (\omega/c)^2\varepsilon_2\mu_2}$ occurs to be equal to $1.88 \cdot 10^6\ cm^{-1}$ . This value is 15.4 times larger than $k_0$ . If $\varepsilon_1 = \varepsilon_3 = 6.5$ would be taken then $h$ would be equal to $1.42 \cdot 10^7\ cm^{-1}$ being 116 times larger now than $k_0$. In the first case $\lambda_{pl} = 33.4\ nm$, $q_{1,2} \approx 1.9 \cdot 10^6\ cm^{-1}$, in the second case $\lambda_{pl} = 4.44\ nm$, $q_{1,2} \approx 1.4 \cdot 10^7\ cm^{-1}$. A Gaussian beam waist dimension would be equal to $3\lambda_{pl} \times (\lambda_{pl} / \pi + d) = 100\ nm \times 12\ nm$ in the 1st case and 13.5 nm × 2.4 nm in the 2nd case.

On seeing these results a possibility to produce a subwave length focusing seems to be achievable. However the Gaussian beam in the device under the discussion has great dimensions being measured in the plasmon wave length. It is of 30-100 wave length in cross section and 300-1000 wave length long along the axis. The question how high would the losses appears.

Propagation losses will be defined by means of the Equ. (9) with a complex dielectric constant $\varepsilon_2 = \varepsilon_2' + i\varepsilon_2''$ in it. The result of the calculations is the following. At $\lambda_0 = 514.6\ nm$, $\varepsilon_1 = \varepsilon_3 = 1$, $\varepsilon_2 = \varepsilon_2' + i\varepsilon_2'' = -10.67 + i0.3$ /9/ the plasmon decaying path $\lambda_{pl} \cdot \operatorname{Re} h / \operatorname{Im} h$ that is the length at which the amplitude of a plasmon decreases by $e^{-1}$ factor is 35.5 wave lengths. If $\varepsilon_1 = \varepsilon_3 = 6.5$ then the plasmon decaying path is 26 wave lengths. If $\varepsilon_1 = \varepsilon_3 = 10$ then the plasmon wave length is 273.5 times shorter than the vacuum wave length and the plasmon decaying path is 8.2 wave lengths.

Great propagation losses become a serious problem in applications. However there is a possibility to use amplifying adjacent layers $\varepsilon_{1,3} = \varepsilon_{1,3}' - i\varepsilon_{1,3}''$ for outweighing losses. There are reports on successful experiments of this kind /10/. The calculations show that with $\varepsilon_1 = 9 - i0.27$ it occurs $h = (147.4 + i5.3 \cdot 10^{-3})(\omega_0 / c)$. For preventing amplification of a s-plasmon of small slowing-down the amplifying layer should be taken thin.

The issue of losses is a key one if high power pulses would be of interest. It is a pulse of the amplification saturation intensity that could go through a plasmon wave guide with an amplifying layer. The saturation intensity $I_s$ is equal to $\hbar\omega_0 / \sigma\tau$. If the amplification cross section $\sigma$ would be $2.5 \cdot 10^{-20}\ cm^2$ and the relaxation time $\tau$ would be $10^{-10}\ s$ then $I_s = 1.5 \cdot 10^{11}\ watt / cm^2$ at $\lambda_0 = 514.6\ nm$. It is a moderately high value of the intensity. If the pulses 100 fs would be taken then the pulse duration should be taken instead of $\tau$. The value $I_s = 1.5 \cdot 10^{14}\ watt / cm^2$ results. However such an intensity and even notably higher could be made without subwave length focusing.

The issue of the saturation intensity does not arise if a plasmon wave guide would be used in a receiver of a threshold sensitivity. It is a device of space selection that should be used in such a receiver. The device could be in form of two lenses with coinciding foci and a diaphragm in the common focus. The other option of the device contains a nanocylinder cut-off, see Fig. 5. A noise power in a device with an amplifying layer of a threshold sensitivity with an amplifying layer should be investigated thoroughly.

Now let the 2nd device, a resonance optical antenna by Pohl be investigated. In the Pohl device there is an Au strip 40 nm thick ~ 260 nm long and 45 nm wide with a gap ~ 20 nm wide in the middle. While studying this device a question of what is a degree that the optical antenna overlaps with the focal spot $\lambda_0$ in diameter. In other words it is a question of an antenna scattering cross section.

First it will be a check what are the dimensions of such a strip in resonance. The calculations will be made in an approximate manner by changing the strip by a cylinder cut and by using an appropriate characteristic equation /4/. At $a = 21.13\ nm$ the calculated resonance length occurs to be $L = 244.6\ nm$. This value is well consistent with ~ 260 nm from the paper.

A field strength $\vec{E}$ in the gap between the antenna halves is equal to the electric displacement $\vec{D}$ in the substance of the antenna. This statement arises as a result of using a boundary condition on the equality of normal components of the electric displacement on both sides of the boundary between media with different dielectric constants.

For the simplicity sake the metal strip will be exchanged by a metal particle of the volume $V$ in an external field $E_0$ of a frequency $\omega_0$. The particle has $\varepsilon_1(\omega) = \varepsilon_1'(\omega) + i\varepsilon_1''(\omega)$ and presents by itself a spheroid in the field which electric component is directed along the most long principal axis. The formulas from the paper /12/ where the nanoparticle resonance is discussed will be used. The total dipole momentum induced in the particle is described by the formula

$$P(\omega_0) = \chi(\omega_0)V\left[ E_0 + i\frac{2k_0^3}{3}P(\omega_0) \right]. \tag{10}$$

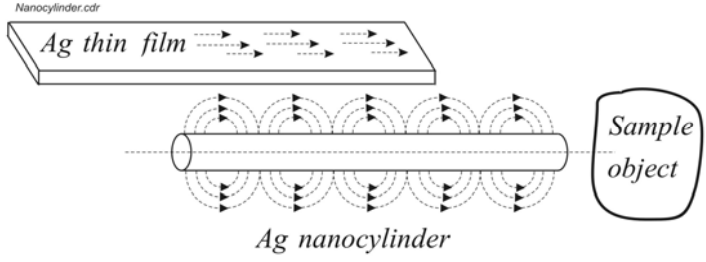

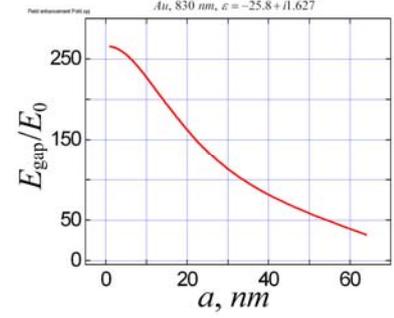

Fig. 5.                                                Fig. 6.

Here $\chi(\omega)$ is the diagonal element of the spheroid susceptibility tensor. At $V/\lambda^3 << 1$ the $\chi(\omega)$ value could be calculated in the quasistatic approximation. The second member in (10) accounts of energy losses due to emission of radiation by introducing of a radiation field reaction $\vec{E}_r = (2/3)k_0^3 \vec{P}(\omega)$, $k_0 = \omega_0/c$. The field $\vec{E}_r$ is selected in the way that is common in the classic field theory /13/: the work produced by $\vec{E}_r$ over the dipole $\vec{P}$ is equal to the emitted energy. Solving (10) relative $P(\omega_0)$ gives

$$P(\omega_0) = \frac{\chi(\omega_0)}{1 - i\dfrac{2k_0^3}{3}\chi(\omega_0)V}VE_0 .$$
(11)

The correction member in the denominator of (11) has a strong dependence on the frequency due to $k_0^3$ and $\chi(\omega_0)$. The later value becomes very large at plasmon resonances in the nanoparticle. The next formula exists for the susceptibility of a ellipsoidal particle:

$$\chi(\omega) = \frac{1}{4\pi}\frac{\varepsilon_1(\omega) - 1}{2 - [1 - \varepsilon_1(\omega)]A} .$$
(12)

The depolarization factor $0 < A < 1$ characterizes the eccentricity of the particle.

For Au the $\varepsilon_1^{''}$ value is small in the visible region. Therefore the peak value of $\chi(\omega_0)$ follows on maximizing (12) relative $\varepsilon_1^{'}(\omega)$. On $\varepsilon_1^{''}, V/\lambda^3 << 1$ the maximum arises at $2 - [1 - \varepsilon_1^{'}(\omega_{res})]A \approx 0$. It is seen that the radiative attenuation should be taken into consideration in the case when $(1 - \varepsilon_1^{'}) \times (4\pi^2 V/3\lambda^3) \approx \varepsilon_1^{''}A$.

Putting the resonance $\varepsilon_1^{'}$ defined by the condition $2 - [1 - \varepsilon_1^{'}(\omega_{res})]A \approx 0$ into (11) gives

$$P_{res} = -i\frac{3}{2k_0^3}\frac{E_0}{\dfrac{3\lambda^3}{(2\pi)^2 V}\dfrac{\varepsilon_1^{''}A}{\varepsilon_1 - 1} - 1} .$$
(13)

The external field is directed along the most long spheroid axis.

Let the field in the gap be calculated. It is equal to $\vec{E}_{gap} = \vec{D} = \vec{E}_0(1 + 4\pi P_{res}/V)$.

$$\left|\frac{\vec{E}_{gap}}{\vec{E}}\right| \approx \left(\frac{\varepsilon_1^{''}A}{(\varepsilon_1^{'} - 1)} - \frac{4\pi^2 a^3 f}{3\lambda^3}\right)^{-1} , \quad f = \frac{L}{2a} .$$
(14)

The parameters of the experiment in /3/ are: $a = 21.12\ nm$, $L = 244.6\ nm$, $A = 0.091$, $f = 5.791$, $\lambda_0 = 830\ nm$, $\varepsilon_1 = -25.8 + i1.627$ /9/. In the result the value $\left|\vec{E}_{gap} / \vec{E}\right| = 168.8$ arises. The dependence of $\left|\vec{E}_{gap} / \vec{E}\right|$ on $a$ for Au at the wave length $\lambda_0 = 830\ nm$ is presented in Fig. 6.

According to the definition the cross section $\sigma$ is equal to $\langle W \rangle / (cE_0^2 / 8\pi)$. Here $\langle W \rangle$ is a mean value of the dipole radiative power, $\langle W \rangle = ck_0^4 d_0^2 / 2$ /13/, $d_0$ is the amplitude of the dipole oscillations that is equal to $P_{res}$ in the case under the consideration.

$$\langle W \rangle = \frac{3cE_0^2\lambda^2}{16\pi^2[3\varepsilon_1'' \lambda^3 A/(2\pi)^2 V(\varepsilon_1 - 1) - 1]^2}.\tag{15}$$

Therefore

$$\frac{\lambda^2}{\sigma} = \left[\frac{2\sqrt{\pi}\lambda^3\varepsilon_1'' A}{(2\pi)^2 V(\varepsilon_1 - 1)} - \frac{2\sqrt{\pi}}{3}\right]^2,\tag{16}$$

$A = \int_0^\infty a^2 c(s + c^2)^{-3/2}(s + a^2)^{-1}ds$, $f = c / a$ - the eccentricity.

The dependence of the ratio $\lambda_0^2 / \sigma$ for Au nanowires in vacuum on the nanowire radius $a$ at $\lambda_0 = 830\ nm$ is shown in Fig. 7. At $a = 21.12\ nm$ the ratio $\lambda_0^2 / \sigma$ is equal to ~ 40. At $a = 128\ nm$ the ratio $\lambda_0^2 / \sigma$ is equal to ~ 2.7.

The conclusion follows from the above consideration. The Yablonovitch device as well as the Pohl device both able to localize an electromagnetic field on a spot that would be much smaller than $\lambda^2$ and both are subject to great losses.

Let the improvement that suits for the Yablonovitch device as well as the Pohl device be discussed. Let it be a resonance atom (quantum dot) placed near a thin film or spheroid. Being exited the atom would radiate and excite some nanodevice. The atom absorption cross section is equal to $\cong \lambda_0^2 \Delta\omega_{rad} / \Delta\omega_{inh}$. Here $\Delta\omega_{rad}$ is a radiative width and $\Delta\omega_{inh}$ is a Doppler width in the case of an atom or a phonon mediated width for a quantum dot. In the visible spectrum for an atom $\Delta\omega_{rad} / \Delta\omega_{inh} \approx 0.01$, for a quantum dot the $\Delta\omega_{rad} / \Delta\omega_{inh}$ value could reach the 0.1 value.

The atom near the spheroid, see Fig. 8, is in the field of two modes of the inhomogeneous space. The first mode is a spherical mode $\vec{n}_{10}$, the exiting wave. The second mode is a $TM_0$ resonance surface plasmon of the nanospheroid. The falling wave excites the atom and practically does not excite the $TM_0$ plasmon. The spontaneous radiative transition of the atom happens mainly with the emission of a photon into the $TM_0$ plasmon. The efficiency of the $TM_0$ plasmon excitation is 100% as efficient.

The ratio of the probability of emission into the plasmon to the probability of emission in a free space is equal to

$$F = E_{pl}^2 \rho_{pl} / E_{free}^2 \rho_{free}.\tag{17}$$

Here $E_{pl}$, $E_{free}$ are the fields of the corresponding field mode, that are normalized to have a single photon in the mode, $\rho_{pl}$, $\rho_{free}$ - the state densities of the corresponding field /14/. Calculations show that at the maximal link between the atom and the plasmon

$$F \approx (h/k_0)^3 \varepsilon_1^2 Q \,.$$ 

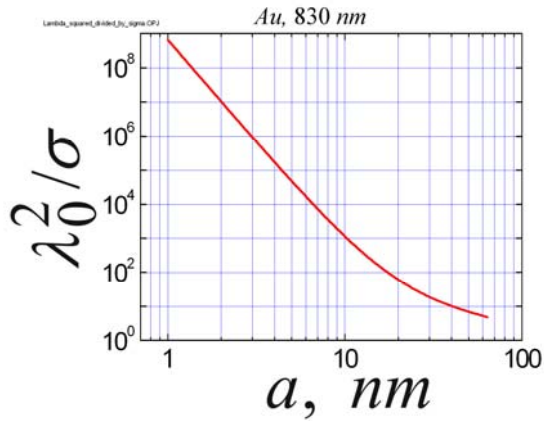

Fig. 7.

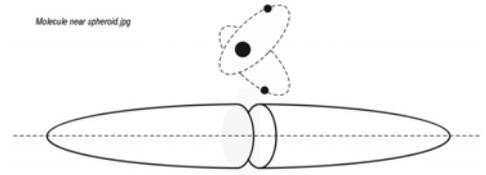

Fig. 8.

Here $h$ - the plasmon wave number, $Q$ - the quality factor of the plasmon resonance in the nanospheroid of the length $L = \pi/h$, $k_0 = \omega/c$. For the nanospheroid $2\ nm$ in diameter the ratio $h/k_0 \approx 33$ and $Q \approx 70$ (due to losses in the substance of the nanospheroid, radiative losses are small) for $\lambda_0 = 514.6\ nm$, $\varepsilon_1 = -10 + i0.3$ ( $Ag$ ). For such meanings of values in $F$ it is equal to $3 \cdot 10^8$. Indeed the exited atom emits radiation in the plasmon mainly. The acquired value of $F$ should be recalculated without use of technique of the approximation theory. However it will stay extremely large even in this case.

When $F$ is large then $\Delta\omega_{inh}$ becomes large also. Now $\Delta\omega_{inh}$ is equal to the inverse time of radiation into a plasmon that is to $\Delta\omega_{rad} F$. The link with the plasmon should be decreased till a moderately small value. This could be accomplished by placing the atom at an enlarged distance from the nanospheroid. By choosing $F$ to be equal to $2 \div 3$ it is possible to get the 100% efficiency of the emission of the radiation into the plasmon practically without the increase in the $\Delta\omega_{rad}/\Delta\omega_{inh}$ ratio.

In conclusion, two types of devices of subwave light focusing are considered. The 1st one presents by itself a thin film of a well reflecting metal which a converging beam of surface plasmons with a wave number $h >> \omega_0/c$ is excited on. The waist of the beam has the cross section that is much smaller than $\lambda_0^2$. The 2nd device is a linear antenna of nanodimensions with a gap in the middle. A resonance plasmon is exited in the antenna. The field in the gap concentrates on a spot much less than $\lambda_0^2$. In the 1st case as well as in the 2nd case a great intensification of the field in the focusing spot appears. However the total power in the output spot is decreased due to propagation losses in the 1st case and due to a small value of the antenna excitation cross section in the 2nd case. The effect of the film or antenna excitation could be increased by the excitation of the adjacent atom or quantum dot.